\documentclass[%
 reprint,
 superscriptaddress,
 amsmath,
 amssymb,
 aps,
 longbibliography,
 prl
]{revtex4-2}

\usepackage{graphicx}% Include figure files
\usepackage{dcolumn}% Align table columns on decimal point
\usepackage{bm}% bold math
\usepackage{hyperref}% add hypertext 
\usepackage{amssymb}
\usepackage{amsmath,bm}
\usepackage{float}
\usepackage[usenames,dvipsnames]{color}
\usepackage[export]{adjustbox}
\usepackage{gensymb}
\usepackage{soul,xcolor}
\usepackage{multirow}
\usepackage{tikz}
\usepackage{dcolumn}

\newcommand{\ket}[1]{\left|  #1 \right\rangle}

\begin{document}

\title{An entanglement protocol to measure atomic parity violation at sub 0.1\% precision}
 
 \author{Diana P. L. Aude Craik}
\affiliation{Institute for Quantum Electronics, Department of Physics, Eidgenössische Technische Hochschule Z\"{u}rich, Otto-Stern-Weg 1, 8093 Zurich, Switzerland}%

\begin{abstract}
This paper proposes a scheme to measure atomic parity violation (APV) in barium ions at $\leq0.1\%$ precision. The scheme is based on using multi-ion entangled states to common-mode reject parity-conserving systematic shifts and selectively detect a parity-violating vector light shift. This measurement protocol eliminates the need to suppress a leading systematic by 11 orders of magnitude, as is required in the single-ion measurement scheme \cite{Fortson93}. Furthermore, the protocol can be combined with the use of integrated photonic waveguides in an architecture that is scalable to large ion numbers, with a proportional increase in measurement precision.
\end{abstract}

\maketitle

%%%%% ---- Main text start ---- %%%%
Atomic parity violation (APV) is a uniquely sensitive probe of beyond-Standard-Model (BSM) physics and of our understanding of the electroweak force. It arises from the parity-violating weak interaction between the electrons and the atomic nucleus, mediated by the $Z^0$ boson. The interaction leads to mixing between energy levels of opposite parity in an atom, weakly allowing transitions that are strictly forbidden by electromagnetism. 
 A measurement of such a parity-violating transition amplitude allows one to determine the charge of the nucleus under the weak force, $Q_W$. This, in turn, translates into a measurement of a fundamental parameter of the electroweak sector of the Standard Model (SM): the electroweak mixing angle (or Weinberg angle), $\theta_w$ \cite{Weinberg1976}. 

\begin{figure}
\centering
\includegraphics[width=\linewidth]{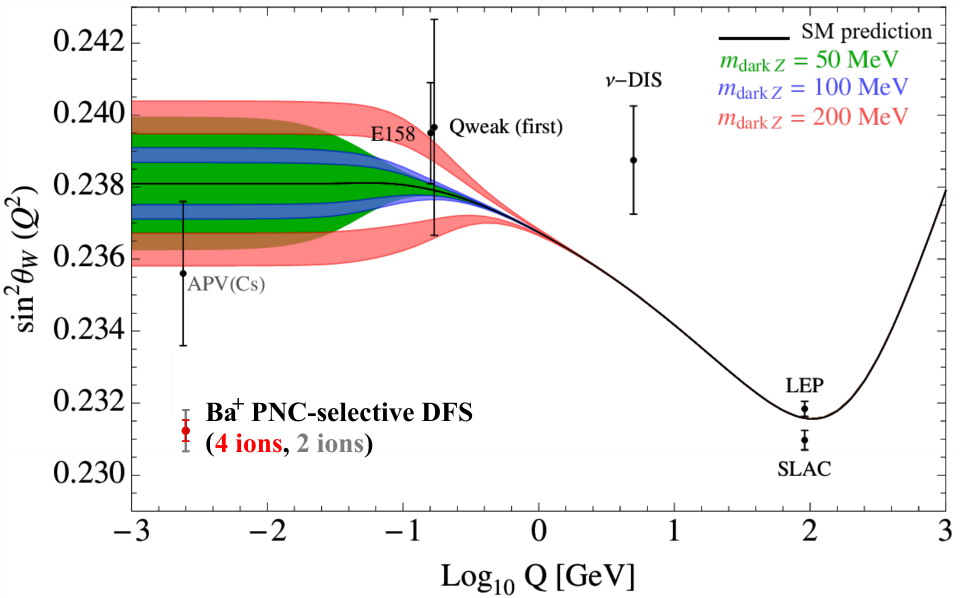}
\caption{\label{fig:sensitivity}\small Running of $\text{sin}^2(\theta_w)$, as a function of interaction momentum transfer $Q$ (Figure modified from \cite{Davoudiasl_2014}). The black curve shows the SM prediction, and black datapoints indicate previous measurements. Shaded regions exemplify how BSM scenarios -- in this case, new Z bosons of different masses, from \cite{Davoudiasl_2014} -- can cause deviations from the SM prediction. The projected experimental error bar for the measurement scheme proposed here is shown on the bottom left in red (4-ion scheme) and grey (2-ion scheme). Note that the projection point is plotted at the relevant momentum transfer for Ba$^+$ of $Q\sim 2.5$\,MeV/c, but is displaced along the y-axis for clarity.\vspace{-2em}}
\end{figure}

The way $\theta_w$ varies (or ``runs") as a function of interaction momentum transfer is precisely predicted by the SM -- any measurement that can detect potential deviations from the SM prediction is therefore an excellent probe for new physics \cite{arcadi2019newphysicsprobesatomic} (see Fig.~\ref{fig:sensitivity}). Collider and neutrino experiments have probed $\theta_w$ at energies ranging from 100\,MeV to 100\,GeV and, thus far, have found agreement with the SM, with the Large Electron-Positron Collider (LEP) having made the most precise measurement to date. However, at energies below 10\,MeV, a single datapoint constrains the running of $\theta_w$: a measurement of APV in cesium made in 1997 \cite{Cs1997}. Although APV experiments probe transitions that are only at the eV energy scale, the achievable measurement precision is such that the measurement is sensitive to miniscule virtual contributions of much heavier particles, in the TeV mass range. At 0.35\% precision, the Cs measurement probed new physics up to the 20\,TeV scale and was, until recently, the leading constraint on several BSM scenarios \cite{wieman2019atomic, Safronova2018, PDG}.

 Given the exceptional discovery potential of APV, the past 30 years have seen ongoing efforts to make new measurements in several atomic and molecular species, including barium, cesium, ytterbium, radium and francium \cite{Fortson1993,radiumAPV, Tsigutkin2009, YbAPV,  APVradioactive, YaleAPV, ParisAPV, versolato2011, APVFrancium, newCsAPV, Sherman2005, dutta2016, DijckThesis, DijckBaPaper, Koerber2003}. However, no experiment has thus far produced a measurement that matches the constraining power of the 1997 Cs data point. This is largely due to the challenging systematic effects that hamper traditional measurement schemes. In the scheme proposed in 1993 for the measurement of APV in a single barium ion \cite{Fortson93}, for instance, residual circular polarization can induce a spin-dependent quadrupole shift that mimics that parity-violating signal but couples $10^7$ times more strongly to the relevant energy levels \cite{Koerber2003}. To sufficiently suppress this systematic would require the alignment and stabilization of three angles (misalignment angles of the analyzing magnetic field, and relative phase and axis of standing wave (SW) fields interacting with the ion) to $10^{-4}$ radian precision \cite{Koerber2003}, a severely challenging task. 
 
\begin{figure*}
  \centering
    \includegraphics[width=0.9\textwidth]{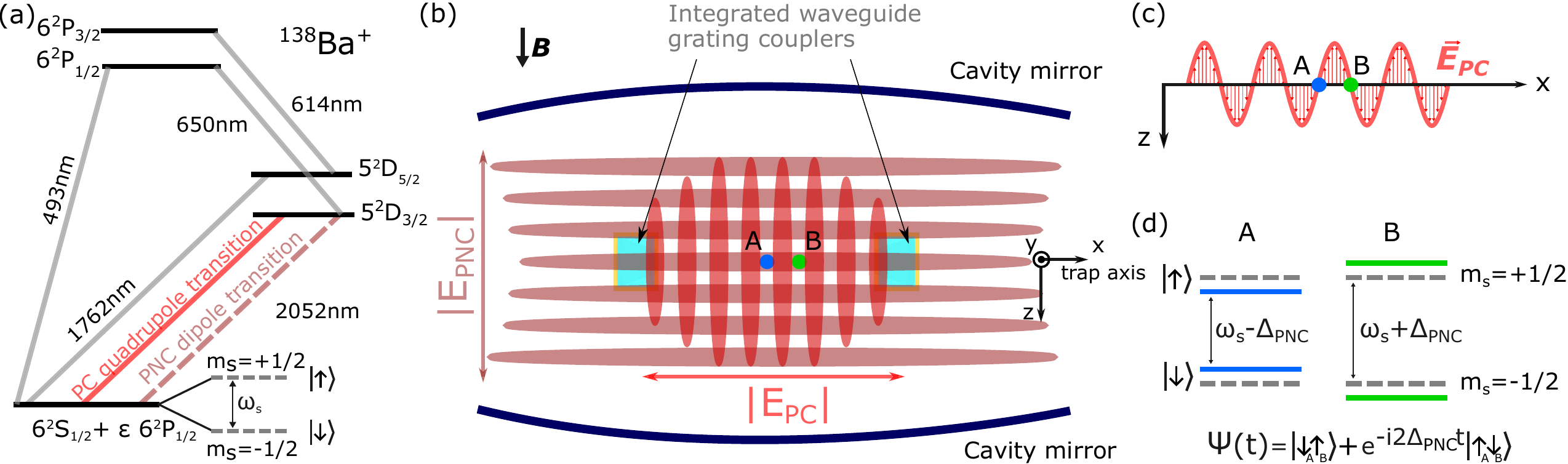}
    \caption{\small (a) The parity-violating interaction of the electrons with the nuclear weak charge introduces a miniscule admixture, $\varepsilon \sim 10^{-11}$, of $P_{1/2}$ into the $S_{1/2}$ ground state and there is hence a very weakly allowed parity non-conserving (PNC) dipole transition between $S_{1/2}$ and $D_{3/2}$. Any observable obtained by driving this PNC transition alone would be quadratically suppressed by $\varepsilon$ \cite{JShermanThesis, Koerber2003}. Instead, Fortson proposed measuring an interference effect, which is linear in $\varepsilon$, between the PNC transition and the parity conserving (PC) electric quadrupole transition between the same two energy levels. (b) Diagram of the crossed standing waves (SWs) experimental setup proposed here, with the antinodes of the SWs indicated in red. The $\mathbf{E_{PNC}}$ SW, which drives the PNC dipole transition, is formed along the $z$-axis by the cavity. The $\mathbf{E_{PC}}$ wave, which drives the PC quadrupole transition, is generated by integrated waveguide grating couplers along the trap axis ($x$ axis). Two ions, A and B, are placed at an antinode of $\mathbf{E_{PNC}}$ and at nodes of $\mathbf{E_{PC}}$. (c) Amplitude plot of $\mathbf{E_{PC}}$ vector field along the trap axis, with ions A and B at successive nodes (arrows indicate that the field is linearly polarized along the $z$-axis). (d) Since ion B experiences a $\pi$ phase shift between the two SWs relative to ion A (and hence an effective flip in parity), its levels are shifted in the opposite direction by the parity-violating vector light shift, $\mathbf{\Delta_\text{PNC}}$. When the ions are placed in an entangled DFS state, $\ket{\psi(t)} = \frac{1}{\sqrt{2}}\ket{\uparrow_A\downarrow_B}+e^{-i2\Delta_{\text{PNC}}t}\ket{\downarrow_A\uparrow_B}$, where $\ket{\uparrow}$ and $\ket{\downarrow}$ are the $m_s=-\frac{1}{2}$ and $m_s=+\frac{1}{2}$ levels of the ground state, 
    the energy difference between the two parts of the superposition, given here by 2$\Delta_{\text{PNC}}$, appears as a relative phase that can be extracted by Ramsey spectroscopy. This relative phase will remain unaffected in the presence of any systematic shifts, as long as they are common-mode to ion A and B. Since parity-conserving systematic shifts do not change sign under a parity transformation, they will be common mode to both ions and factor out as an unimportant global phase in the DFS state. \vspace{-1em}
    } 
    \label{fig:MeasurementScheme}
\end{figure*}
 This paper proposes a new protocol to overcome this problem by making the measurement in a different way. The protocol employs a multi-ion entangled state to isolate the parity violating signal, rejecting any (parity conserving) systematic shifts that are common to both ions. I refer to this scheme henceforth as a parity-non-conserving-selective decoherence-free subspace, or a \emph{PNC-selective DFS}. Due to the differential nature of the scheme, even with percent-level misalignments, an anticipated measurement precision of 0.05\% or 0.1\% respectively would be achieved for a 4-ion and 2-ion PNC-selective DFS (see Fig.~\ref{fig:sensitivity}). Furthermore, when used in combination with state-of-the-art integrated photonic waveguides to generate passively-stable optical standing waves, this measurement scheme offers an architecture that is scalable to larger numbers of ions, with a proportional increase in measurement precision.

 \textbf{Measurement scheme: a PNC-selective DFS}. In 1993, N. Fortson proposed that one can measure APV in a single $\text{Ba}^+$ ion by configuring two perpendicular standing waves (SWs) of 2052nm light, $\mathbf{E_{PC}}$ and $\mathbf{E_{PNC}}$, resonant with the electric quadrupole transition from $S_{1/2}$ to $D_{3/2}$ such that the ion sits at the node of $\mathbf{E_{PC}}$ and the antinode of $\mathbf{E_{PNC}}$ \cite{Fortson93} (see Fig.~\ref{fig:MeasurementScheme}). The nodal wave, $\mathbf{E_{PC}}$, drives a parity-conserving (PC) electric quadrupole transition (since it is the gradient of the electric field that couples to the quadrupole transition). The antinodal wave, $\mathbf{E_{PNC}}$, drives a parity-non-conserving (PNC) electric dipole transition between the $S_{1/2}$ ground state and $D_{3/2}$ (see Fig.~\ref{fig:MeasurementScheme}a). A vector light shift arises on the ground $S_{1/2}$ state due to interference between these two transition amplitudes. 

 This parity-violating light shift is spin dependent and acts, on the ground state, much like an ``effective magnetic field", shifting the energies of the $m_s=-\frac{1}{2}$ and $m_s=+\frac{1}{2}$ Zeeman sublevels in opposite directions (see Fig.~\ref{fig:MeasurementScheme}d). This produces an overall vector shift, $\mathbf{\Delta_\text{PNC}}$, in the Larmor precession frequency of the $6S_{1/2}$ ground state, given by %Equation~\ref{eq:PNCshift}). 
  \begin{equation}
 \label{eq:PNCshift}
     \mathbf{\Delta_\text{PNC}} = \eta\langle2(\mathbf{E_{PNC}\cdot\nabla})\mathbf{\dot{E}_{PC}}+\mathbf{E_{PNC}}\times (\nabla \times \mathbf{\dot{E}_{PC}})\rangle_t
 \end{equation} 
 where the $t$ subscript denotes a time average, overdots indicate time derivatives and $\eta$ is a T-even pseudoscalar of magnitude $\approx 10^{-15}ea_0\omega/\Omega^\text{quad}$, where $\omega/\Omega^\text{quad}$ is the ratio of the optical frequency of the SWs to the Rabi frequency of the quadrupole transition \cite{Fortson93}. When a strong laser field ($E_\text{PNC}\approx 1.5\times10^6$\,V/m) is used to drive the PNC transition, $|\mathbf{\Delta_\text{PNC}}|/{2\pi}$ can be made as large as $\approx 0.4$\,Hz \cite{JShermanThesis}.

Unlike any parity conserving shift, $\mathbf{\Delta_\text{PNC}}$ will reverse sign under a parity transformation. A parity transformation requires flipping all directions in space: from the point of view of the ion, this is equivalent to simply changing the phase relationship between the two SWs by $\pi$. Therefore, if we can measure the differential light shift experienced by two ions interacting with the same light fields $\mathbf{E_{PC}}$ and $\mathbf{E_{PNC}}$, but with one ion seeing a $\pi$ shift in relative phase between them, then we will isolate the PNC shift, since all other parity-conserving shifts will be common-mode rejected in the measurement. 

Indeed, Fortson points out in \cite{Fortson93} that the phasing of the SWs can be adjusted to discriminate $\mathbf{\Delta_\text{PNC}}$ from competing systematics, and later Ref.~\cite{JShermanThesis} briefly speculates about the possibility of making parallel measurements of several ions placed in parity-reversed positions, but neither consider the use of entangled states known as decoherence-free subspaces (DFS), likely because such techniques have only much more recently begun to be regularly applied in quantum metrology \cite{RoosDFS, Kotler_2014, Manovitz2019, wilzewski2024}.

 With the emergence of such techniques, Ref.~\cite{PVDFS2010} proposed a measurement scheme employing a two-ion DFS created by entangling an ion experiencing Fortson's SW geometry with another ion experiencing no light fields. This rejects common-mode background noise (such as magnetic field fluctuations), but does not reject the leading PNC-mimicking spin-dependent quadrupole shift systematic, which would hence still need to be independently suppressed.
 
 The scheme proposed here involves co-trapping two ions on successive nodes of the $\mathbf{E_{PC}}$ field, as shown in Fig.~\ref{fig:MeasurementScheme}b,c. Relative to ion A, ion B sees a $\pi$ phase shift between $\mathbf{E_{PC}}$ and $\mathbf{E_{PNC}}$. By applying an entangling gate (using, for instance, the protocol in \cite{Sawyer_2021}), we can then create the entangled state $\ket{\Psi(t)}=\ket{\downarrow\uparrow}+e^{-i2\Delta_\text{PNC}t}\ket{\uparrow\downarrow}$, where $\ket{\downarrow}$ and $\ket{\uparrow}$ correspond to the $m_s=-1/2$ and $m_s=+1/2$ Zeeman sublevels of the $S_{1/2}$ ground state and $\Delta_\text{PNC}$ is the shift in Larmor frequency experienced by each ion due to the PNC transition amplitude. The situation is depicted in Fig.~\ref{fig:MeasurementScheme}d: because the two ions experience Larmor shifts of opposite sign, the relative phase between the $\ket{\downarrow\uparrow}$ and $\ket{\uparrow\downarrow}$ parts of $\ket{\Psi}$ evolves at twice the PNC shift frequency, doubling the signal size and creating, in effect, a PNC-selective DFS. Two-ion Ramsey spectroscopy \cite{RoosDFS, bollinger1996optimal, RamseyMethod} can then be used to extract this relative superposition phase, giving a direct measurement of $\Delta_\text{PNC}$.
 
 The improvement in the feasibility of the APV measurement that is afforded by the scheme proposed here is considerable: in the single-ion version originally proposed, a small amount of residual circular polarization in the $\mathbf{E_{PNC}}$ field would produce a spin-dependent quadrupole shift that couples $10^7$ times more strongly to the ground state than the PNC shift we wish to measure \cite{Koerber2003, Fortson1993}. If the PNC shift is to be measured at 0.1\% precision, this systematic must hence be suppressed by 11 orders of magnitude. This quadrupole shift is indistinguishable from the PNC shift, except in how it transforms under parity, which is what the scheme exploits -- the spurious quadrupole shift, as well as any other parity-conserving systematic shift, has the same sign for ions A and B. It hence factors out as an unimportant global phase in the superposition, which does not affect the Ramsey measurement.
 
\textbf{Phase swap to cancel differential systematics.} The cancellation of systematic shifts relies on the environment seen by the two ions being the same (except for the $\pi$ phase shift between $\mathbf{E_{PNC}}$ and $\mathbf{E_{PC}}$). In any real experiment, however, the two ions will experience a difference in other parameters caused by spatial gradients in the trap, such as a difference in static magnetic field, trap RF fields, and stray electric fields. To suppress systematic shifts that could be caused by a difference in environments between ions A and B, the phase of $\mathbf{E_{PNC}}$ can be translated by $\pi$ and the measurement can then be repeated -- this effectively swaps the phase relationship between $\mathbf{E_{PNC}}$ and $\mathbf{E_{PC}}$ seen by ions A and B, leading to a change in sign of $\Delta_\text{PNC}$ and hence of the relative phase in the superposition state $\ket{\Psi(t)}$. The averaged difference of measurements in the two phase configurations would hence produce (to the extent that the phase translation is perfect) a systematics-free measure of $\Delta_\text{PNC}$.

\textbf{Standing wave generation and phase stabilization.} As indicated in Fig.~\ref{fig:MeasurementScheme}b, the $\mathbf{E_{PC}}$ SW can be generated by interfering two beams emanating from waveguide gratings registered to the chip's surface \cite{gillenapaper,Ricci2023} and fed by integrated photonic waveguides. Ion position relative to SWs generated in this way has recently been shown to be passively stable to $1.6$\,nm from shot-to-shot, greatly facilitating single-ion calibration measurements of the SW \cite{Ricci2023}. Furthermore, this architecture is easily scalable to multiple SW zones, to which large numbers of entangled ions can be distributed.

The ions can be positioned in $\mathbf{E_{PC}}$ by tuning the trap's DC electrode potentials and hence no differential phase control is needed for the two beams that interfere to produce this SW: i.e. they can be fed from a single input integrated waveguide that is split on chip to deliver light to the two gratings shown in Fig.~\ref{fig:MeasurementScheme}b. 

In contrast, the phase of $\mathbf{E_{PNC}}$ must be tunable. The antinode of this SW must be overlapped with the trap axis in initial alignment, and the phase of this SW must be shifted by $\pi$ when the phase-swap procedure described above is performed. Phase control of $\mathbf{E_{PNC}}$ can either be implemented by interfering free-space beams \cite{MainzSW, oxfordSW} 
 or creating this SW using a free-space beam and an in-vacuum optical cavity (Fig.~\ref{fig:MeasurementScheme}b) as, for instance, in \cite{Cetina2013}. Alternatively, an integrated waveguide solution could also be implemented using on-chip phase shifters \cite{GrangeLiN2022, hogle2023high}. While attractive due to its scalability \cite{Mordini2025,kwon2024multi} and positional stability, the latter approach needs careful analysis due to the presence of a non-zero running wave amplitude at the anti-node of the grating-generated SW (for the $\mathbf{E_{PC}}$ SW this is not a problem, since the ions sit at nodes of this SW, where the running wave amplitude is zero).

To stabilize the phase in the cavity design, we would wish to lock the cavity length in such a way that it would remain locked when $\mathbf{E_{PNC}}$ is pulsed on and off since, to map out SW intensity in calibration sequences, we would drive Rabi oscillations with the SW light (as in \cite{Ricci2023}). This can be implemented by locking the cavity length to a transfer laser that does not interact with the atom \cite{CetinaThesis, JonSimonThesis}. Phase control and stabilization of $\mathbf{E_{PNC}}$ in the integrated waveguide design can be accomplished with an on-chip Mach-Zender interferometer actuated, for instance, by lithium niobate phase shifters \cite{GrangeLiN2022, GrangeLiN2024}. 

 \textbf{Scaling up the measurement protocol.} By using an $N$-ion maximally correlated entangled state, we can increase the signal-to-noise ratio by a factor of $N$. Moving to a four-ion state of the form $\ket{\Psi_4(t)}=\ket{\downarrow\uparrow\downarrow\uparrow}+e^{i8\Delta_\text{PNC}t}\ket{\uparrow\downarrow\uparrow\downarrow}$ (using, for instance an entanglement protocol such as in \cite{Insbruck14qubitMSgate}), will octuple the PNC signal as compared to the single-ion experiment (from 400\,mHz to 3\,Hz). The successful implementation of the pilot experiment proposed here will pave the way to scaling up further -- a future measurement of the PNC light shift at 1\,mHz precision on a 14-ion state would amount to a fractional precision of $\sim0.01$\%, probing BSM physics up to the $\sim150$\,TeV energy scale.

\textbf{Isotope-chain measurements to reduce sensitivity to theory uncertainty.} An electronic structure coefficient is required to translate an APV measurement to a bound on the weak charge of the nucleus, $Q_W$ \cite{Safronova2018}. This coefficient is currently thought to be largely isotope independent and is therefore expected to cancel in ratios formed from APV measurements in different isotopes, up to a dependence on the neutron charge distribution (``neutron skin") \cite{neutronSkinAPVBrown2009,neutronskinISAPV1999}. Recent work suggests that the theory uncertainties on the ``neutron skin" are highly correlated between isotopes and that the overall theoretical uncertainty on isotopic ratios of APV can be reduced to 0.2\% \cite{neutronSkinAPVBrown2009}. The measurement protocol proposed here can be applied to several isotopes of barium (there are five spinless, stable isotopes) to produce an isotope-chain measurement, as recently done in ytterbium \cite{YbAPV}, but at increased precision. Such a measurement would enhance sensitivity to BSM electron-proton interactions \cite{neutronskinISAPV1999, neutronSkinAPVBrown2009} and provide a useful benchmark for theoretical models of the neutron skin, thought to be the leading remaining source of theoretical uncertainty in isotopic APV ratios.

\textbf{Conclusion.} The novelty of the measurement protocol proposed here is in the fact that it rejects parity conserving spin-dependent shifts, removing the requirement to suppress a leading systematic by 11 orders of magnitude. This brings the previously exceptionally challenging goal of measuring APV at 0.1\% precision well within reach.

\begin{acknowledgments}
\textbf{Acknowledgments.} The author thanks Jonathan Home, Alfredo Ricci Vásquez and Gillen Beck for useful discussions on integrated optics technology; Jeremy Flannery, Roland Matt and Luca I. Huber for helpful exchanges on Ramsey spectroscopy systematics and entangling-gate implementations, and Daniel Kienzler for thoughtful comments on the manuscript. The author thanks the SNSF for approving an SNSF Starting Grant to fund the implementation of this APV measurement scheme in barium ions (grant number 225809, approved in the SNSF Starting Grants 2024 round). This research has received funding from the European Union's Horizon 2020 research and innovation programme under the Marie Skłodowska-Curie Grant Agreement No. 795121, and from the Army Research Office, under Cooperative Agreement Number W911NF-23-2-0216.

\end{acknowledgments}

\bibliography{references}
%%%%% ---- End of main text ---- %%%%%%
\end{document}